\pdfoutput=1
\documentclass[sigconf]{acmart}

\AtBeginDocument{%
  \providecommand\BibTeX{{%
    \normalfont B\kern-0.5em{\scshape i\kern-0.25em b}\kern-0.8em\TeX}}}

\usepackage[utf8]{inputenc}
\usepackage[T1]{fontenc}
\usepackage{url}
\usepackage[capitalise]{cleveref}
\usepackage{amsmath,amsfonts,amsthm,commath,dsfont,nicefrac,xfrac}
\usepackage{mathtools}
\usepackage{bm}
\usepackage{xspace}
\usepackage{xcolor}
\usepackage{booktabs,makecell,multirow}
\usepackage{tikz}
\usepackage{pgfplots}
\usepackage{subcaption}
\usetikzlibrary{patterns}
\usetikzlibrary{calc,positioning,fit}
\usepgfplotslibrary{groupplots,units}
\usepackage{enumitem}

\newlength\figureheight 
\newlength\figurewidth 
\setlist[enumerate]{leftmargin=20pt}
\setlist[itemize]{leftmargin=20pt}

\crefname{section}{§}{§§}
\Crefname{section}{§}{§§}

\usepackage[skip=3pt]{caption}
\pgfplotsset{
  compat=1.9,
  unit code/.code 2 args={\si{#1#2}} 
}

\newcommand{\system}[0]{{AutoIndex}\xspace}

\newcommand{\alex}{{\tt ALEX}\xspace}
\newcommand{\psql}{{\tt Postgres}\xspace}
\newcommand{\rocksdb}{{\tt RocksDB}\xspace}

\newcommand{\rmi}{{\sf RMI}\xspace}
\newcommand{\pgm}{{\sf PGM}\xspace}
\newcommand{\lmdb}{{\sf LMDB}\xspace}

\newcommand{\sta}{{\sf Tall}\xspace}
\newcommand{\stb}{{\sf Wide}\xspace}
\newcommand{\enva}{{\sf ShortRTT}\xspace}
\newcommand{\envb}{{\sf LongRTT}\xspace}

\definecolor{barOrange}{HTML}{bb5555}
\definecolor{BlueColor}{HTML}{bb5555}
\definecolor{RedColor}{HTML}{bb5555}
\definecolor{OrangeColor}{HTML}{bb5555}
\definecolor{barYellow}{HTML}{ee9944}
\definecolor{barLightGreen}{HTML}{A3A847}
\definecolor{barGreen}{HTML}{5588bb}
\definecolor{GreenColor}{HTML}{5588bb}

\definecolor{vintageblack}{HTML}{484043}
\definecolor{vintageyellow}{HTML}{FFC805}
\definecolor{vintagegreen}{HTML}{38A528}
\definecolor{vintageorange}{HTML}{FF4D25}
\definecolor{vintagepurple}{HTML}{AE56E2}

\begin{document}

\title[Automatically Finding Optimal Index Structure]{Automatically Finding Optimal Index Structure}
\subtitle{Extended Abstracts}

\author{Supawit Chockchowwat,
Wenjie Liu,
Yongjoo Park}
\email{{supawit2,wenjie3,yongjoo}@illinois.edu}
\affiliation{%
  \country{CreateLab @UIUC, USA}
}








\renewcommand{\shortauthors}{Supawit Chockchowwat, Wenjie Liu, and Yongjoo Park}

\begin{abstract}
Existing \emph{learned indexes} (e.g., RMI, ALEX, PGM) optimize 
the \emph{internal regressor} of each node,
\emph{not the overall structure} such as index height, the size of each layer, etc.
In this paper, we share our recent findings that 
\textbf{we can achieve significantly faster lookup speed
by optimizing the structure as well as internal regressors}.
Specifically, our approach (called \system) expresses \emph{the end-to-end lookup time
as a novel objective function},
and searches for optimal design decisions using a purpose-built optimizer.
In our experiments with state-of-the-art methods, 
\system achieves 3.3$\times$--7.7$\times$ faster lookup for the data stored on local SSD,
and 1.4$\times$--3.0$\times$ faster lookup for the data on Azure Cloud Storage.
\end{abstract}

\maketitle

\section{Introduction}
\label{sec:intro}

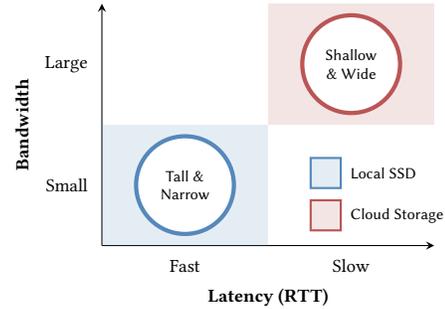
\begin{figure}[t]
\pgfplotsset{
mystyle/.style={
    axis lines = left,
    width=60mm,
    height=48mm,
    legend style={
        at={(-0.2,1.1)},anchor=south west,column sep=2pt,
        draw=black,fill=none,line width=.5pt,
        /tikz/every even column/.append style={column sep=5pt},
        font=\footnotesize,
    },
    tick label style={font=\footnotesize},
    label style={font=\footnotesize\bf},
    xtick style={draw=none},
    ytick style={draw=none},
    xmin=1,
    xmax=5,
    xtick={1,2,3,4,5},
    xticklabels={,Fast,,Slow,},
    xlabel=Latency (RTT),
    ymin=1,
    ymax=5,
    ytick={1,2,3,4,5},
    yticklabels={,Small,,Large,},
    ylabel=Bandwidth,
    clip=false,
}
}

\tikzset{
mynode/.style={
    circle,draw=gray!50!white,minimum width=13mm,ultra thick,
    align=center,font=\scriptsize\sf,fill=white,
},
myshade/.style={
    fill opacity=0.15,inner xsep=0,inner ysep=0
},
}

\centering
\begin{tikzpicture}

\begin{axis}[
    mystyle,
]

\node[fit={(axis cs: 1,1) (axis cs: 3,3)},myshade,fill=GreenColor] {};
\node[fit={(axis cs: 3,3) (axis cs: 5,5)},myshade,fill=BlueColor] {};

\node[mynode,draw=GreenColor] at (axis cs: 2,2) {Tall \& \\Narrow};
\node[mynode,draw=BlueColor] at (axis cs: 4,4) {Shallow \\ \& Wide};

\def\x{3.5}
\node[myshade,fill=GreenColor,anchor=west,minimum width=4mm,minimum height=4mm,
    draw=GreenColor,thick] 
    (L1) at (axis cs: \x, 2.2) {};
\node[anchor=west,font=\scriptsize\sf] at ($(L1.east)+(0.1,0)$) {Local SSD};
    
\node[myshade,fill=BlueColor,anchor=west,minimum width=4mm,minimum height=4mm,
    draw=BlueColor,thick] 
    (L2) at (axis cs: \x, 1.5) {};
\node[anchor=west,font=\scriptsize\sf] at ($(L2.east)+(0.1,0)$) {Cloud Storage};

\end{axis}

\end{tikzpicture}
\caption{
Expected optimal structures by I/O characteristics. 
Our approach, \system, can automatically find optimal index structures 
as well as internal regressors.}
\label{fig:intro}
\end{figure}

Indexes enable fast lookup and are employed
by many data systems for fast search and analytics:
search engines~\cite{elasticsearch,splunk1}, key-value stores~\cite{leveldb,rocksdb,raju2017pebblesdb}, RDBMS~\cite{momjian2001postgresql,mysql}, etc.
Conventional indexes (e.g., B-tress~\cite{aguilera2008practical,wu2010efficient,sowell2012minuet}, red-black trees~\cite{bayer1972symmetric, guibas1978dichromatic, besa2013concurrent}, skip lists~\cite{zhang2019s3}) 
offer $O(\log n)$ lookup speed for $n$ data items.
Learned indexes~\cite{kraska2018case,ding2020alex,tang2020xindex,wang2020sindex,ferragina2020pgm,galakatos2019fiting,kipf2020radixspline} 
can offer even lower latencies
based on more compact representations~\cite{FERRAGINA2021107}.
To achieve high performance,
indexes are typically optimized specifically for target storage media
(e.g., SSD~\cite{li2010tree}, NVMe~\cite{wang2020pa},
memory~\cite{binna2018hot}, L1 cache~\cite{rao2000making}). 
For example, a B-tree node fits in a disk page (e.g., 4KB).
RMI~\cite{kraska2018case} has three layers with the fast random-access assumption.

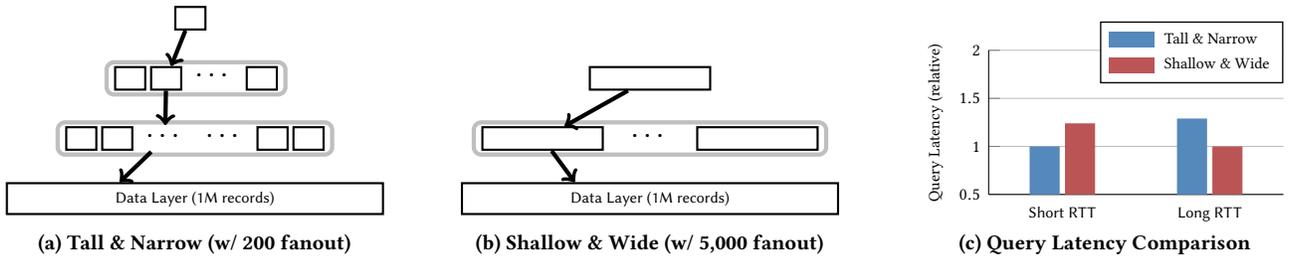
\begin{figure*}[t]

\tikzset{
mynode/.style={
    draw=black,thick,font=\scriptsize\sf,
    minimum height=3mm,minimum width=4mm,
},
mylarge/.style={
    mynode,minimum width=16mm,
},
mylayer/.style={
    draw=gray!50!white,ultra thick,rounded corners=1mm,
    inner ysep=0.5mm,inner xsep=1mm
}
}

\pgfplotsset{mystyle/.style={
    ybar,
    xtick=data,
    width=55mm,
    height=35mm,
    bar width=4mm,
    ymin=0.5,
    ymax=2.0,
    axis y line*=none,
    axis x line*=none,
    xmin = 0.5,
    xmax = 2.5,
    xtick={1,2},
    xticklabels = {\textsf{Short RTT}, \textsf{Long RTT}},
    xtick style={draw=none},
    ytick={0, 0.5, 1.0, 1.5, 2.0},
    tick label style={font=\scriptsize},
    legend style={
        at={(1.0,1.2)},anchor=north east,column sep=2pt,
        draw=black,fill=white,line width=.5pt,
        /tikz/every even column/.append style={column sep=5pt},
        font=\scriptsize,
    },
    legend cell align={left},
    legend columns=1,
    ylabel near ticks,
    xticklabel shift=-1mm,
    label style={font=\scriptsize},
    ylabel={Query Latency (relative)},
    ymajorgrids,
    area legend,
    legend image code/.code={%
        \draw[#1, draw=none] (0cm,-0.1cm) rectangle (0.6cm,0.1cm);
    },
    clip=false,
}}

\begin{subfigure}[b]{0.32\textwidth}
\centering
\begin{tikzpicture}

\def\h{0.8}
\def\d{0.05}

\node[mynode] (L1) at (0,0) {};

\node[mynode] (L21) at ($(L1)+(-0.8,-\h)$) {};
\node[mynode,anchor=west] (L22) at ($(L21.east)+(\d,0)$) {};
\node[mynode,anchor=west,draw opacity=0,font=\Large\bf] (L23) at ($(L22.east)+(\d,0)$) {$\cdots$};
\node[mynode,anchor=west] (L24) at ($(L23.east)+(\d,0)$) {};

\node[mynode] (L31) at ($(L21)+(-0.65,-\h)$) {};
\node[mynode,anchor=west] (L32) at ($(L31.east)+(\d,0)$) {};
\node[mynode,anchor=west,draw opacity=0,font=\Large\bf] (L33) at ($(L32.east)+(\d,0)$) {$\cdots$};
\node[mynode,anchor=west,draw opacity=0,font=\Large\bf] (L34) at ($(L33.east)+(\d,0)$) {$\cdots$};
\node[mynode,anchor=west] (L35) at ($(L34.east)+(\d,0)$) {};
\node[mynode,anchor=west] (L36) at ($(L35.east)+(\d,0)$) {};

\node[mynode,minimum width=50mm] (D) at ($(L31)!0.5!(L36) + (0,-\h)$) {Data Layer (1M records)};

\node[mylayer,fit={(L21.north west) (L24.south east)}] {};

\node[mylayer,fit={(L31.north west) (L36.south east)}] {};

\draw[->,draw=black,ultra thick] (L1) -- (L22);
\draw[->,draw=black,ultra thick] (L22) -- (L33);
\draw[->,draw=black,ultra thick] (L33) -- ($(D.north)+(-1,0)$);

\end{tikzpicture}
\caption{Tall \& Narrow (w/ 200 fanout)}
\label{fig:motivation:a}
\end{subfigure}
\hfill
\begin{subfigure}[b]{0.32\textwidth}
\centering
\begin{tikzpicture}

\def\h{0.8}
\def\d{-0.2}

\node[mylarge] (L1) at (0,0) {};

\node[mylarge,draw opacity=0,font=\Large\bf] (L22) at ($(L1)+(0,-\h)$) {$\cdots$};
\node[mylarge,anchor=east] (L21) at ($(L22.west)+(-\d,0)$) {};
\node[mylarge,anchor=west] (L23) at ($(L22.east)+(\d,0)$) {};

\node[mynode,minimum width=50mm] (D) at ($(L21)!0.5!(L23) + (0,-\h)$) {Data Layer (1M records)};

\node[mylayer,fit={(L21.north west) (L23.south east)}] {};

\draw[->,draw=black,ultra thick] (L1) -- (L21);
\draw[->,draw=black,ultra thick] (L21) -- ($(D.north)+(-1,0)$);

\end{tikzpicture}
\caption{Shallow \& Wide (w/ 5,000 fanout)}
\label{fig:motivation:b}
\end{subfigure}
\hfill
\begin{subfigure}[b]{0.32\textwidth}
\centering
\begin{tikzpicture}
\begin{axis}[
    mystyle
]

\addplot[fill=barGreen,draw=none]
table[x=x,y=y] {
x y
1 1
2 1.29
};

\addplot[fill=barOrange,draw=none]
table[x=x,y=y] {
x y
1 1.24
2 1.0
};

\addlegendentry{Tall \& Narrow}
\addlegendentry{Shallow \& Wide}

\end{axis}
\end{tikzpicture}
\vspace{-4mm}
\caption{Query Latency Comparison}
\label{fig:motivation:c}
\end{subfigure}

\caption{A motivating example for optimizing the entire index structure.
Two candidate structures in (a) and (b). (a) has a lower branching factor (200);
thus, it is quicker to retrieve each node while the index is higher.
(b) has a greater branching factor (5,000); thus, its height is lower while retrieving each node takes more time.
Figure (c) shows that 
depending on system environments (i.e., \textsf{Short RTT} vs.~\textsf{Fast RTT}),
different structures can show better performance (note: the latency of ``High and Narrow'' is used as a unit latency).
See the text for the simulation setup.
}
\label{fig:motivation}
\end{figure*}

%
%
%
%
%
%

\paragraph{Problem}

\emph{The performance of these indexes is suboptimal} when
they operate in an environment different from what they are designed for.
We have observed this suboptimality (thus, missed opportunity)
as we develop a cloud-optimized document store~\cite{chockchowwat2021airphant}.
Also, see \cref{sec:motivation} for a concrete example with B-tree indexes.


\paragraph{Opportunity \& Challenge}

We observe that we can greatly improve lookup speed
by carefully choosing structural parameters
such as the number of layers, layer sizes, 
the types and accuracy of internal regressors, etc.
Our intuition is as follows.
As depicted in \cref{fig:intro},
we should prefer shallow and wide indexes 
if a storage device has very long latency (round-trip time),
because we want to reduce the number of I/O operations.
In contrast, if the latency is very small in comparison to bandwidth,
tall indexes should be preferred.
However, finding an optimal design is non-trivial in practice
because there are exponentially many candidates (\cref{sec:method:params}).


\paragraph{Our Approach}

We propose a new index builder (called \system) that
\emph{can learn the optimal structure
in a principled manner
by navigating in a high-dimensional design space.}
Specifically, we formulate a novel optimization problem
that expresses an end-to-end lookup time in terms of
core design parameters 
(e.g., the number of layers, the size of each node, regressors, etc.)
as well as I/O performance.
By solving the optimization problem
using our highly parallel search method,
we can build a low-latency index for large-scale data.

Note that our approach is \emph{orthogonal to learned indexes}~\cite{kraska2018case,ding2020alex,tang2020xindex,wang2020sindex,ferragina2020pgm,galakatos2019fiting,kipf2020radixspline}, which propose the use of regressors for branching functions 
(instead of exact pointers).
In contrast, \emph{this work focuses on optimizing the overall index structure}.
This work is in line with
    the recent advances
        at the intersection of data systems and machine learning~\cite{anderson2013brainwash, park2016visualization, park2017database, marcus2018nashdb, kraska2019sagedb, li2019qtune, naru2019, park2020quicksel, park2019blinkml, park2018verdictdb, bater2020saqe, park2017active}.





\section{Motivation \& Scope}
\label{sec:motivation}


\subsection{Motivation}
\label{sec:dominant}

Our intuition is the following:
the structure of an index is \emph{crucial} for its performance
because depending on 
the system environment it operates on,
the optimal index structure can \emph{vastly differ}.
For example, optimal index structures can be different
if we store data on network-attached devices instead of local SSD,
\emph{even if we index the same dataset.}
We illustrate this with a concrete example.

\paragraph{Concrete Example}

Suppose two different B-tree structures: \sta and \stb. (Note: our model formulation in \cref{sec:method} supports more than B-trees.) \sta consists of 4 KB nodes, where each node has 200 fanout. \stb consists of 100 KB nodes, where each node has 5,000 fanout. Both \sta and \stb index the same dataset with one million distinct keys. The dataset is stored in 4 KB pages.

To index the dataset,
\sta needs three layers (because the third layer can hold up to $200^3 = 8M$ pointers).
Likewise, \stb needs two layers (because the second layer can hold up to $5000^2 = 25M$ pointers). 
Note that while \stb is shallower than \sta, fetching each node of \stb takes longer because its page size is 25$\times$ larger.
\cref{fig:motivation:a,fig:motivation:b} depict these structures.

Interestingly, neither of these two indexes (i.e., \sta and \stb)
is superior to the other;
that is, there is no single dominant index structure
that offers faster lookup speed in all different system environments.
To illustrate this, we suppose two storage devices, \enva and \envb.
\enva operates with 5 ms latency (RTT) and 100 MB/sec bandwidth.
\envb operates with 100 ms latency (RTT) and 100 MB/sec bandwidth.
That is, their bandwidth is the same, 
but their latencies are different. (Note that our formulation in \cref{sec:method} handles more general cases.)

Now, we show that (i) \sta offers higher performance than \stb if we store data on \enva,
and that (ii) \stb offers higher performance than \sta if we store data on \envb.
For this,
we adopt a widely used formula: (data transfer time) $=$
(latency) + (data size) / (bandwidth).
For instance, fetching a single page of the dataset stored on \enva
takes 5 ms + 4 KB / (100 KB/ms) = 5.04 ms.

\vspace{2mm}
\textbf{Environment---\enva: \stb is 24\% slower than \sta.}
\begin{itemize}
\item \sta needs 3 nodes and 1 data page = 
3 $\times$ (5ms + 4KB / (100KB/ms)) + (5ms + 4KB / (100KB/ms)) = 20.16 ms
\item \stb needs 2 nodes and 1 data page =
2 $\times$ (5ms + 500KB / (100KB/ms)) + (5ms + 4KB / (100KB/ms)) = 25.04 ms
\end{itemize}

\vspace{2mm}
\textbf{Environment---\envb: \sta is 29\% slower than \stb.}
\begin{itemize}
\item \sta needs 3 nodes and 1 data page = 
3 $\times$ (100ms + 4KB / (100KB/ms)) + (100ms + 4KB / (100KB/ms)) = 400.16 ms
\item \stb needs 2 nodes and 1 data page =
2 $\times$ (100ms + 500KB / (100KB/ms)) + (100ms + 4KB / (100KB/ms)) = 310.04 ms
\end{itemize}

\cref{fig:motivation:c} summarizes this relative performance strength.
For each environment, the figure reports the \emph{relative} difference in 
end-to-end lookup time.
It proves that different index structures offer higher lookup
performance, depending on the storage device.

\paragraph{Need for Efficient Algorithm} 

While we use a relatively simple example to illustrate the core intuition,
finding an optimal index structure is non-trivial to solve by hand,
because there are exponentially many configurations we need to consider.
To efficiently find an optimal index, 
    we develop an intelligent learning technique
    that considers the entire index building 
        as an optimization problem.

\subsection{Scope of This Work}
\label{sec:scope}

\paragraph{Read-only Index}

We focus on building read-only indexes on a sorted key-value dataset.
There are two reasons.
First, this is a novel attempt at optimizing the \emph{entire} index
by casting it into a statistical optimization problem. Thus, we aim to ensure
at least we can build best-in-class read-only indexes.
Second, our long term plan is to integrate our index
with write-oriented data systems based on LSM trees~\cite{o1996log}.
Note that segments of an LSM tree are not updatable; they are only
merged occasionally to compose bigger (next level) segments. 
For these compaction operations, it suffices to build read-only
indexes as part of compaction.

\paragraph{Point Lookup}

We focus our evaluation on point lookup---fetching a value associated with
a search key.
There are two reasons.
First, by optimizing point lookup, we also optimize range search because
they involve essentially the same internal operations---traversing
nodes from top to bottom.
Second, range search can easily be supported by extending point lookup.
To retrieve values for a key range \texttt{(begin, end)},
we can find the offset for \texttt{begin} using point lookup, then
continue fetching data until we see \texttt{end}.

\paragraph{Persistent Index}

We focus on the index we persist on storage (e.g., SSD, NVMe, flash drive), 
not the index maintained in main memory.
This is because \textbf{we aim to enable faster lookup against large-scale data}
kept in persistent databases (e.g., RocksDB, PostgreSQL, SQLite, etc).
For this reason, our cost analysis (for fetching data) must incorporate
storage-specific data transfer speed.

\section{Index as Statistical Optimization}
\label{sec:method}

We build an index by solving a statistical optimization problem.
Specifically,
our algorithm builds the fastest index
for the data stored on a certain storage system,
where the storage system is abstracted by
the \emph{data transfer function} $T(o)$.
$T(o)$ is the time it takes to retrieve the data $o$
(serialized on a storage device; see \cref{sec:objective} for its usage).
In the rest of this section,
we formulate an optimization problem and discuss how to solve it.

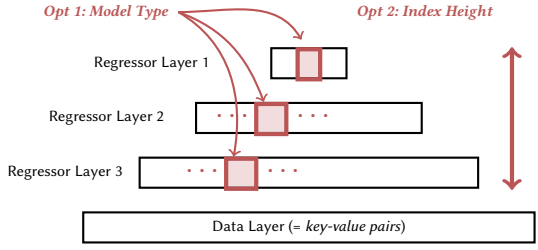
\begin{figure}[t]
  \centering
  
\tikzset{
mynode/.style={
    anchor=north,draw=black,thick,font=\scriptsize,
    minimum height=4mm,minimum width=16mm,
}
}

\begin{tikzpicture}

\coordinate (T) at (0,0);


\node[mynode,minimum width=10mm] (L1) at (T) {};

\node[mynode,minimum width=30mm] (L2) at ($(L1.south) + (0,-0.3)$) {};

\node[mynode,minimum width=45mm] (L3) at ($(L2.south) + (0,-0.3)$) {};

\node[mynode,minimum width=60mm,font=\sf\scriptsize,inner ysep=1mm] (L4) 
    at ($(L3.south) + (0,-0.3)$) 
    {Data Layer (= \emph{key-value pairs})};

\node[anchor=east,font=\sf\scriptsize] at ($(L1.west)+(-0.7,0)$) {Regressor Layer 1};
\node[anchor=east,font=\sf\scriptsize] at ($(L2.west)+(-0.3,0)$) {Regressor Layer 2};
\node[anchor=east,font=\sf\scriptsize] at ($(L3.west)+(-0.1,0)$) {Regressor Layer 3};

\node[
    mynode,ultra thick,minimum width=3mm,draw=RedColor,
    fill=RedColor,fill opacity=0.2
] 
(M1) at ($(L1.north)+(-0,0)$) {};

\node[
    mynode,ultra thick,minimum width=4mm,draw=RedColor,
    fill=RedColor,fill opacity=0.2
] 
(M) at ($(L2.north)+(-0.5,0)$) {};
\node[anchor=east,text=RedColor,font=\Large,inner xsep=0] 
at ($(M.west) + (-0,0)$) {\textbf{$\cdots$}};
\node[anchor=west,text=RedColor,font=\Large,inner xsep=0] 
at ($(M.east) + (0.1,0)$) {\textbf{$\cdots$}};

\node[
    mynode,ultra thick,minimum width=4mm,draw=RedColor,
    fill=RedColor,fill opacity=0.2
] 
(M2) at ($(L3.north)+(-0.9,0)$) {};
\node[anchor=east,text=RedColor,font=\Large,inner xsep=0] 
at ($(M2.west) + (-0,0)$) {\textbf{$\cdots$}};
\node[anchor=west,text=RedColor,font=\Large,inner xsep=0] 
at ($(M2.east) + (0.1,0)$) {\textbf{$\cdots$}};

\node[
    mynode,ultra thick,minimum width=3mm,
    draw=none,
] 
(O1) at ($(T)+(-2.7,0.7)$) {\color{RedColor} \textit{\textbf{Opt 1: Model Type}}};
\draw[->,thick,draw=RedColor] (O1.east) to[out=0,in=120] ($(M1.north)+(-0.1,0.1)$);
\draw[->,thick,draw=RedColor] (O1.east) to[out=-20,in=110] ($(M.north)+(-0.1,0)$);
\draw[->,thick,draw=RedColor] (O1.east) to[out=-30,in=80] ($(M2.north)+(-0.1,0)$);

\coordinate (B) at (2.7,0);
\draw[<->,ultra thick,draw=BlueColor] (B) -- ($(B)+(0,-1.9)$);

\node[
    mynode,ultra thick,minimum width=3mm,
    draw=none,
    anchor=north west
] 
(O2) at ($(T)+(0.5,0.7)$) {\color{BlueColor} \textit{\textbf{Opt 2: Index Height}}};

%

%

\end{tikzpicture}  
  
\vspace{2mm}
\caption{Index structure and optimization parameters.}
\label{fig:model_design}
\end{figure}

\subsection{Index Class}
\label{sec:method:class}

An \emph{index class} is a class of indexes we can express. 
The more general an index class is, the larger 
the set of indexes that we can represent; however, finding an optimal index can be more computationally expensive.
\textbf{Our index class generalizes hierarchical indexes} 
such as B-trees~\cite{bayer1972organization,comer1979ubiquitous}
and learned indexes~\cite{kraska2018case}, as described below.

An \emph{index} consists of layers (see \cref{fig:model_design}).
There are two types of layers: \emph{regressor layers} and \emph{the data layer}. 
The data layer contains records sorted by keys.
Each regressor layer consists of regressor(s).
A regressor outputs a range \texttt{(start, end)}
which is the data range (in byte offsets) we need to fetch in the next layer.
If the next layer is a regressor layer, we fetch regressor(s); 
if the next layer is a data layer, we fetch key-value pairs (including the search key).

\paragraph{Querying} First, we fetch the entire root layer (i.e., Regressor Layer 1) containing multiple regressors.
We use an appropriate regressor (for the given search key) to obtain an offset range for the next layer.
This step continues until we reach the data layer.
Note that we fetch the entire range at a time;
that is,
if there are $L$ regressor layers, we fetch $L+1$ times until we obtain a desired key-value.


\subsection{Parameter Space}
\label{sec:method:params}

A \emph{parameter tuple} $\Theta$ captures the design considerations for an index.
The goal of our statistical optimization is to find the optimal $\Theta$ ($= \Theta^*$) 
that minimizes the end-to-end lookup time $\ell$.
$\Theta$ consists of the following parameters:
\begin{enumerate}
\item Number $L$ of regressor layers ($R_1, \ldots, R_L$). 
    (Note: For convenience, we use $R_{L+1}$ for the data layer.)
\item Regressor Type ($\bm{C} = C_1, \ldots, C_L$): 
A regressor layer $R_l$ consists of regressor(s) of the same type (e.g., linear regression, step functions),
which we denote by $C_l$.
We search for the optimal regressor type for each regressor layer.
\item Precision ($\bm{\lambda} = \lambda_1, \ldots, \lambda_L$) of regressor layers:
A regressor outputs a range. While we ensure the range contains an appropriate regressor (in the next index layer) or
a desired key-value pair (in the data layer), we must tune the sizes of those ranges, which we achieve via $\bm{\lambda}$.
\end{enumerate}
In determining $\bm{\lambda}$, there is a natural trade-off. 
That is, if we build an accurate regressor (with a fine $\lambda_i$), its range output tends to be small, but the size
of the regressor itself (and the regressor layer) must be big (e.g., with more coefficients for higher accuracy).

Using $\Theta = (L, \bm{C}, \bm{\lambda})$,
we express the end-to-end lookup latency in terms of $\Theta$ as described in the next section.

\subsection{Objective Function for Latency}
\label{sec:objective}

We express a (tight) upper-bound on
the end-to-end lookup time $\ell$ in terms of $\Theta$ and
the data transfer function, $T(o)$.
Specifically, $\ell$ is expressed in terms of actual 
regressor layers ($R_1, \ldots, R_L$) created as specified by  $\Theta$;
here, the size of $R_l$ depends on $R_{l+1}$ as well as $\Theta$
(because $R_l$ is to obtain the data fetch range for $R_{l+1}$).

\paragraph{Objective Function}

Let $m(\cdot)$ be a process for creating a regressor. 
The created regressor's output range size is expressed by $\varepsilon$.
\begin{align*}
R_l &\coloneqq m(C_l, \lambda_l; R_{l+1}) \\
& \text{where } R_l \text{ fetches } \varepsilon(C_l, \lambda_l; R_{l+1}) \text{ bytes for } R_{l+1}
\end{align*}
For a lookup, we first fetch the entire root layer ($R_1$); 
then, we subsequently fetch 
every range output by the latest obtained regressor until
the data layer.
Thus, $\ell$ can be expressed as:
\begin{equation}
\begin{split}
\ell(L, \bm{C}, \bm{\lambda}) 
= T(&m(C_1, \lambda_1; R_2))
+ T( \varepsilon(C_1, \lambda_l; R_{2}) )
\\ &+ \sum_{l=2}^L T( \varepsilon(C_l, \lambda_l; R_{l+1}) )
\end{split}
\label{eq:objective}
\end{equation}
The goal is to find $\Theta^* = (L^*, \bm{C}^*, \bm{\lambda}^*)$ at which the above expression is minimized.
We present our parameter search algorithm in \cref{sec:build}.

\paragraph{Discussion}

Our objective function is an \emph{upper bound}
because some data may be cached in memory; 
in these cases, no data transfer is needed.
We empirically observed that our approach 
also delivers strong performance 
in partially cached scenarios.

\subsection{Index Build via Parameter Search}
\label{sec:build}

We provide a high-level sketch of our parameter search algorithm.
We omit theoretical justifications and parallelization techniques.
Our optimization is in the form of branch-and-bound 
algorithms---we \emph{branch} out to 
construct the \emph{best} $R_l$ given (i) already constructed $R_{l+1}, \ldots, R_{L+1}$
and (ii) \emph{optimal upper layers} (i.e., $R_1, \ldots, R_{l-1}$); then, we
\emph{bound} by pruning candidates.
Here, the \emph{optimal upper layers} are obtained via a recursive call of our algorithm,
and the \emph{best} $R_l$ is evaluated using \cref{eq:objective}.
There are three features that make our approach logically sound and efficient:
\begin{enumerate}
\item Principled search for $L^*$: 
In designing our algorithm,
one of the greatest technical challenges is 
finding the optimal number of layers. 
Our algorithm does this by asking ``does stacking another layer reduce the total lookup time?''
based on \cref{eq:objective}. This approach makes our algorithm logically sound.
\item Effective enumeration of candidate $\Theta$:
Our algorithm is not amenable to gradient-descent-style algorithms for two reasons.
First, the parameters for $\Theta$ include discrete choices 
(i.e., integer $L$ and regressor type $\bm{C}$).
Second, the regressor creation process $m(\cdot)$ is not always differentiable.
Nevertheless, our search algorithm can find almost optimal $\Theta$
by effectively navigating in the discrete search space.
\item Highly parallel:
While our algorithm examines many different $\Theta$, 
our total search \& build process is 
highly efficient (comparable to existing \emph{learned indexes} building) 
because our algorithm is designed to run in an \emph{embarrassingly parallel} manner.
\end{enumerate}

\section{Experiment}
\label{sec:benchmark}

We share our latest experiment results.
While we continue making further enhancements in our implementation,
our current results are strong,
with the following key points:
\begin{enumerate}
\item \system's learning algorithm exhibits expected behavior
of finding optimal index structure for given system environments
(\cref{sec:exp:learning}).
\item \system's lookup speed outperforms state-of-the-art methods 
on different storage devices
(\cref{sec:exp:faster}).
\end{enumerate}

\noindent
We present each point after explaining our experiment setup.

\subsection{Setup}
\label{sec:exp:setup}

We conduct experiments on Microsoft Azure.
Our experiments use a large-scale benchmark dataset.

\paragraph{System}

For VMs, we use \texttt{Standard\_D8s\_v3} instances (16 vCPUs, 64 GB memory).
For Small RTT, we use OS disks (Premium SSD LRS, 256 GB, P20-2300 IOPS, 150 MBps).
For Large RTT, we use Azure Cloud Storage (StorageV2)~\cite{azureblobstorage}
connected via NFS v3~\cite{nfs}.

\paragraph{Compared Methods}

We compare \system against several state-of-the-art index libraries:

\begin{enumerate}
\item \rmi: RMI~\cite{kraska2018case} is one of the most commonly studied
learned index implementations. 
RMI comes with an accuracy knob---our results compare RMI with the highest accuracy.
\item \pgm: PGM-Index~\cite{ferragina2020pgm} is one of the latest learned indexes.
It offers provable worst-case bounds.
\item \lmdb: Lightning Memory-Mapped Database Manager (\lmdb)
is one of the fastest open-source B-tree libraries.
A benchmark reports that \lmdb is orders-of-magnitude faster than
commonly used indexing systems such as LevelDB and SQLite3~\cite{lmdb-benchmark}.
\end{enumerate}

\paragraph{Dataset \& Queries} 

We use the Facebook dataset included in ``SOSD: A Benchmark for Learned Indexes''~\cite{sosd-neurips}.
This dataset has 200 million keys.
For querying, we randomly pick a key from the dataset and measure the end-to-end time
for retrieving the value associated with the key
(for each method we compare).
To reduce variance, we report the average over twenty iterations.

\subsection{\system Adapts}
\label{sec:exp:learning}

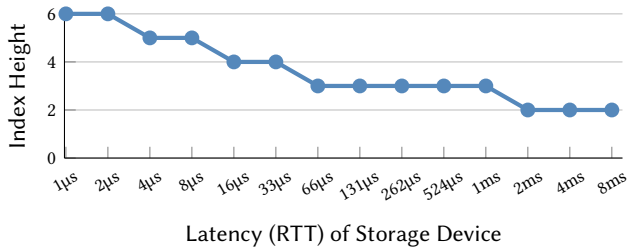
\begin{figure}[t]

\pgfplotsset{mystyle/.style={
    xtick=data,
    width=90mm,
    height=35mm,
    bar width=4mm,
    ymin=0,
    axis y line*=none,
    axis x line*=none,
    xmin = 1000,
    xmax = 10000000,
    xtick={1024, 2048, 4096, 8192, 16384, 32768, 65536, 131072, 262144, 524288, 1048576, 2097152, 4194304, 8388608},
    xticklabels = {1\textmu s, 2\textmu s, 4\textmu s, 8\textmu s, 16\textmu s, 33\textmu s, 66\textmu s, 131\textmu s, 262\textmu s, 524\textmu s, 1ms, 2ms, 4ms, 8ms},
    tick label style={font=\footnotesize},
    xticklabel style={rotate=30},
    legend style={
        at={(-0.2,1.1)},anchor=south west,column sep=2pt,
        draw=black,fill=none,line width=.5pt,
        /tikz/every even column/.append style={column sep=5pt},
        font=\footnotesize,
    },
    legend cell align={left},
    legend columns=4,
    ylabel near ticks,
    label style={font=\normalsize\sf},
    xlabel={Latency (RTT) of Storage Device},
    ylabel={Index Height},
    ymajorgrids,
    area legend,
    legend image code/.code={%
        \draw[#1, draw=none] (0cm,-0.1cm) rectangle (0.6cm,0.1cm);
    },
    clip=false,
}}

\centering
\begin{tikzpicture}
\begin{axis}[
    mystyle,
    xmode=log,
    ymax=6,
]

\addplot[mark=*,ultra thick,draw=GreenColor,
    mark options={fill=GreenColor}
]
table[x=x,y=y] {
x y
1024	6
2048	6
4096	5
8192	5
16384	4
32768	4
65536	3
131072	3
262144	3
524288	3
1048576	3
2097152	2
4194304	2
8388608	2
};


\end{axis}
\end{tikzpicture}

\caption{\system can adapt. 
For greater RTTs, \system properly creates shallower \& wider indexes.}
\label{fig:exp:learning}
\end{figure}

First, we test if \system produces different (and properly optimized) index structures
according to given system environments.
While there are many internal parameters defining the structure of an index,
we specifically examine the number of layers (or equivalently, index height)
because its results are easier to understand than other parameters
(e.g., size of regressor layers, types of regressors).
Expectation: As illustrated in \cref{sec:motivation},
a shallow index should be preferred if the latency (RTT) is large.
Actual: To see if our learning algorithm also exhibits such outcomes,
we run \system for different latency values
while the bandwidth is fixed at 134 MB/sec.
\cref{fig:exp:learning} shows the results.
When the latency is extremely small (i.e., 1 microsecond),
the algorithm tells us that a six-layered index delivers the fastest lookup speed. 
The number of layers gradually decreases (from 6 to 2) as we increase the latency,
which is an expected result.
We compare the actual lookup speed below.

\subsection{\system is Fastest}
\label{sec:exp:faster}

\begin{figure}[t]
\pgfplotsset{mystyle/.style={
    ybar,
    bar shift=0pt,
    xtick=data,
    width=70mm,
    height=38mm,
    bar width=4mm,
    ymin=0,
    axis y line*=none,
    axis x line*=none,
    xmin = 0.5,
    xmax = 4.5,
    xtick={1,2,...,4},
    xticklabels = {\textsf{LMDB}, \textsf{RMI}, \textsf{PGM}, 
        \textbf{\system} 
    },
    xtick style={draw=none},
    tick label style={font=\footnotesize},
    legend style={
        at={(-0.2,1.1)},anchor=south west,column sep=2pt,
        draw=black,fill=none,line width=.5pt,
        /tikz/every even column/.append style={column sep=5pt},
        font=\footnotesize,
    },
    legend cell align={left},
    legend columns=4,
    ylabel near ticks,
    xticklabel shift=-1mm,
    label style={font=\small\sf},
    ylabel={Query Latency},
    ymajorgrids,
    area legend,
    legend image code/.code={%
        \draw[#1, draw=none] (0cm,-0.1cm) rectangle (0.6cm,0.1cm);
    },
    clip=false,
}}

\begin{subfigure}[b]{\columnwidth}
\centering
\begin{tikzpicture}
\begin{axis}[
    mystyle,
    ymax=30,
    ytick={0, 10, ..., 30},
    yticklabels={0ms, 10ms, 20ms, 30ms},
]

\addplot[fill=GreenColor,draw=none]
table[x=x,y=y] {
x y
1 9.45844
};

\addplot[fill=GreenColor,draw=none]
table[x=x,y=y] {
x y
2 22.42997
};

\addplot[fill=GreenColor,draw=none]
table[x=x,y=y] {
x y
3 20.08336
};

\addplot[fill=RedColor,draw=none]
table[x=x,y=y] {
x y
4 2.90077
};

\node[
    anchor=north west,draw=black,font=\footnotesize,align=left,
    fill=white
] at (axis cs: 0.5,35) {The lower, the better};

\end{axis}
\end{tikzpicture}
\vspace{-2mm}
\caption{Query Latency under Small RTT (Local Disk)}
\label{fig:exp:fast:a}
\end{subfigure}

\vspace{4mm}
\begin{subfigure}[b]{\columnwidth}
\centering
\begin{tikzpicture}
\begin{axis}[
    mystyle,
    ymax=300,
    ytick={0, 100, ..., 300},
    yticklabels={0ms, 100ms, 200ms, 300ms},
]

\addplot[fill=GreenColor,draw=none]
table[x=x,y=y] {
x y
1 183.30
};

\addplot[fill=GreenColor,draw=none]
table[x=x,y=y] {
x y
2 97.84
};

\addplot[fill=GreenColor,draw=none]
table[x=x,y=y] {
x y
3 211.19
};

\addplot[fill=RedColor,draw=none]
table[x=x,y=y] {
x y
4 69.64
};

\node[
    anchor=north west,draw=black,font=\footnotesize,align=left
] at (axis cs: 0.5,300) {The lower, the better};

\end{axis}
\end{tikzpicture}
\vspace{-2mm}
\caption{Query Latency under Large RTT (Azure Storage)}
\label{fig:exp:fast:b}
\end{subfigure}

\vspace{2mm}
\caption{
\system (ours) delivers significantly faster query latency
compared to other state-of-the-art methods.
This is because \system optimizes the entire index
according to target system environments.
}
\label{fig:exp:fast}
\end{figure}
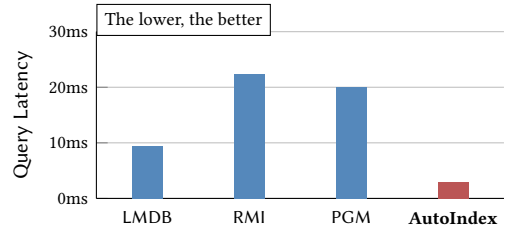
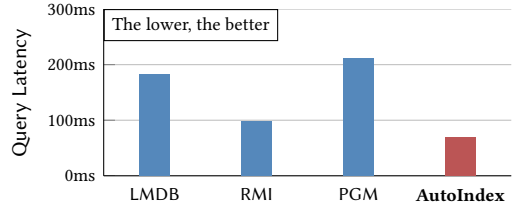

Second, we show that \system outperforms the other indexes.
Our comparison is performed using both local SSDs (small RTT)
and Azure Cloud Storage (large RTT).
\cref{sec:exp:setup} describes 
the details about the methods we compare and also the dataset we use.
\cref{fig:exp:fast} summarizes the results.
\cref{fig:exp:fast:a} compares the lookup time for Small RTT (local disks).
\system is the fastest: it is 3.3$\times$--7.7$\times$ faster than other methods.
\cref{fig:exp:fast:b} compares the lookup time for Large RTT (Azure Storage).
Again, \system is the fastest: it is 1.4$\times$--3.0$\times$ faster than other methods.
This is because \system can optimize an entire index structure
for target system environments.

\section{Ongoing Work}

We are enhancing \system in several orthogonal directions.
First, we are conducting additional theoretical analyses
to formally study the optimality of our parameter search process.
Second, we are extending our optimization problem to incorporate 
possible variance in I/O performance.
Third, we are improving our builder to stream-process very large data.

\begin{acks}
We thank the creators/developers of 
    RMI~\cite{kraska2018case}, PGM-Index~\cite{ferragina2020pgm}, and \lmdb~\cite{lmdb-benchmark},
who open-sourced their software,
    which helped us greatly in conducting accurate experiments.
This work is supported in part by Microsoft Azure.

\end{acks}

\clearpage

\bibliographystyle{ACM-Reference-Format}
\bibliography{references}

\end{document}